\documentclass[twocolumn,showpacs,prb,aps,amsmath,amssymb,superscriptaddress]{revtex4}
\usepackage{amsmath}
\usepackage{graphicx}
\usepackage{rotating}

\usepackage{color}

\def\bbbc{{\mathchoice {\setbox0=\hbox{$\displaystyle\rm C$}\hbox{\hbox
to0pt{\kern0.4\wd0\vrule height0.9\ht0\hss}\box0}}
{\setbox0=\hbox{$\textstyle\rm C$}\hbox{\hbox
to0pt{\kern0.4\wd0\vrule height0.9\ht0\hss}\box0}}
{\setbox0=\hbox{$\scriptstyle\rm C$}\hbox{\hbox
to0pt{\kern0.4\wd0\vrule height0.9\ht0\hss}\box0}}
{\setbox0=\hbox{$\scriptscriptstyle\rm C$}\hbox{\hbox
to0pt{\kern0.4\wd0\vrule height0.9\ht0\hss}\box0}}}}

\definecolor{DarkBlue}{rgb}{0.1,0.1,0.5}
\definecolor{Red}{rgb}{0.9,0.0,0.1}
\definecolor{Green}{rgb}{0.0,0.99,0.0}

\newcommand{\beq}{\begin{eqnarray}}
\newcommand{\bq}{\begin{equation}}
\newcommand{\eeq}{\end{eqnarray}}
\newcommand{\eq}{\end{equation}}

\newcommand{\br}{{\bf r}}

\newcommand{\beqa}{\begin{eqnarray}}
\newcommand{\eeqa}{\end{eqnarray}}

\begin{document}
\title{Wave function for odd frequency superconductors.}
\author{Hari P. Dahal}
\affiliation{Theoretical Division, Los Alamos National Laboratory,
Los Alamos, New Mexico 87545}

\author{E. Abrahams}
\affiliation{Serin Physics Laboratory, Rutgers University, P.O. Box
849, Piscataway, NJ 08855}

\author{D. Mozyrsky}
\affiliation{Theoretical Division, Los Alamos National Laboratory,
Los Alamos, New Mexico 87545}

\author{Y. Tanaka}
\affiliation{Dept. of Applied Physics, Nagoya University,
Chikusa-ku,  Nagoya 464-8603, JAPAN}

\author{A. V.  Balatsky}
\affiliation{Theoretical Division, Los Alamos National Laboratory,
Los Alamos, New Mexico 87545} \affiliation{Center for Integrated
Nanotechnology, Los Alamos National Laboratory, Los Alamos, New
Mexico 87545}\email[] { avb@lanl.gov, http://theory.lanl.gov}

\date{1/15/09}

\begin{abstract}

We revisit the question of nature of odd-frequency
superconductors, first proposed by Berezinskii in 1974.
\cite{berezinskii1974} We start with the notion that order
parameter of odd-frequency superconductors can be thought of as a
time derivative of the odd-time pairing operator. It leads to the
notion
 of the composite boson
condensate.\cite{abrahams1995} To elucidate the nature of
broken symmetry state in odd-frequency superconductors, we consider
a wave function that properly captures the coherent condensate of
composite charge $2e$ bosons in an odd-frequency superconductor. We consider the Hamiltonian which describes the
equal-time composite boson condensation as proposed
     earlier in  Phys.\ Rev.\ B $\textbf{52}$, 1271 (1995). We
      propose a BCS-like wave
     function that describes  a composite condensate  comprised of a
     spin-0 Cooper pair
      and a spin-1 magnon excitation. We derive the
      quasiparticle dispersion, the self-consistent equation
      for the order parameter and the density of states.
We show that the coherent wave function approach recovers all the
known proposerties of odd-frequency superconductors: the
quasi-particle excitations are gapless and the superconducting
transition requires a critical coupling.
\end{abstract}

\maketitle
\section{introduction}
The discussion about possible symmetry types of superconducting order
parameter $\Delta(\textbf{k},\tau)$ ($\tau$ denotes imaginary time) has drawn significant research interest. The conventional
singlet (triplet) superconductor follows $PT
\Delta(\textbf{k},\tau)=\Delta(\textbf{k},\tau)$ $([PT
\Delta(\textbf{k},\tau)=-\Delta(\textbf{k},\tau)])$ under parity $P$
and time $T$ transformations. The singlet ($PT=1$) and  triplet ($PT=-1$) conditions can be satisfied by either taking $P=T=1$ and $P=-1,T=1$ for an even-in-frequency gap,  or
$P=T=-1$ and $P=1,T=-1$ for odd-frequency pairing.

Although mainstream discussions of superconudctivity are for
even-frequency pairing, there is a growing interest in
understanding odd-frequency pairing.
The discussion of unconventional pairing ($P=1,T=-1$) was
initiated by Berezenskii \cite{berezinskii1974} to explain the superfluid phase of $^3$He.
Although his proposal of triplet odd-frequency pairing could
not explain the superfluid phase of $^3$He, it certainly motivated
a search of other possibilities of the pairing symmetries.
Balatsky and Abrahams \cite{balatsky1992} later extended the concept
of odd-frequency pairing  to the singlet superconductor ($P= T=-1$).

Although the realization of the odd-frequency pairing in current
systems is still under debate, several reports consider this
possibility in a a number of systems. Odd-frequency pairing
in the Kondo lattice has been
investigated to study superconductivity in heavy-fermion compounds, \cite{coleman1994}
The proximity effects in a superconductor-ferromagnet
structure\cite{bergeret2001}, a normal-metal/superconductor
 junction\cite{tanaka2007a} and diffusive normal metal/unconventional
superconductor interface\cite{tanaka2007b} have been attributed to odd-frequency
pairing. The $p$-wave singlet odd-frequency pairing is argued to be  a viable pairing in the coexistence region of antiferromagnetism and superconductivity and/or near the
quantum critical point in CeCu$_2$Si$_2$ and CeRhIn$_5$. \cite{fuseya2003} In
addition, hydrated Na$_x$CoO$_2$ is suggested to support an $s$-wave triplet
odd-frequency gap\cite{johannes2004}. Very recently, Kalas {\it et al.} \cite{kalas2008}
 have argued that the boson-fermion cold atom
mixture exhibits $s$-wave triplet odd-frequency pairing above some critical
coupling at which the mixture phase separates.

Motivated by the growing  interest and possibilities of
odd-frequency pairing, here we address the missing part of the
odd-frequency superconductivity discussion: {\em what is the wave
function of the odd-frequency superconductors?} One might wonder how
one can even ask this question given that superconducting
correlations of two fermion operators in odd-frequency
superconductor do not have an equal time expectation value? We
assume (pretty safe assumption in fact) that any state, including
odd-frequency superconductor, does has a many body wave function
that  captures superconducting correlations. Any state of matter has
an associated wave function $|\psi \rangle$ that captures the
amplitude distribution of the particles forming this state. Hence we
are asking exactly this question about the many body wave function
of the odd frequency superconductors. Our wave function builds upon
a long discussion \cite{abrahams95} on the possible order parameter
and {\em equal time} composite operators that capture
superconducting correlations of odd-frequency superconductors in
equal time domain.

We propose a BCS-like pairing wave function for an odd-frequency
superconductor, and study its consequences for the energy
dispersion, superconducting order parameter, and density of states.
The wave function, which describes a condensate of a spin-0 Cooper
pair and a spin-1 magnon excitation, is consistent with the
Hamiltonian suggested earlier in \cite{abrahams1995} to study
 odd-frequency superconductivity. We minimize this Hamiltonian
with respect to the proposed wave function and derive an expression
for the quasiparticle dispersion,
a self consistent gap equation and the density of states. We find that
a) the quasi-particle dispersion is gapless, b) the gap equation has
non zero solution only for a critical value of the coupling, c) the
 density of states is finite even for an  energy
 less than the gap energy, and d) the density of states is
reduced at the gap edge compared to that of the BCS case.

Before introducing the wave function and getting into the details of
the minimization of the Hamiltonian, we would like to show that $PT=1$
can be obtained by taking $P=T=-1$ in $S=0$ singlet case.
Any superconducting order with translational invariance, equilibrium and
broken $U(1)$ symmetry would result in an anomalous (Gor'kov) Green's  function
\begin{equation}
{\cal F}_{\alpha \beta} (\tau,\textbf{k})=\langle T_{\tau} c_{\alpha, \textbf{k}}(\tau) c_{\beta, -\textbf{k}}(0) \rangle,
\end{equation}
where $\alpha, \beta$ are spin indices. We assume that the transition
occurs only in a well defined representation. Thus, for $S=0$  singlet pairing,
we may define
\begin{equation}
F(\tau,\textbf{k}) = \epsilon_{\alpha \beta} {\cal F}_{\alpha \beta}(\tau,\textbf{k}),
\end{equation}
and for $S=1$ triplet pairing,
\begin{equation}
\vec{F}(\tau,\textbf{k}) = (i\widehat{\sigma} \vec{\sigma})_{\alpha \beta} {\cal F}_{\alpha \beta}(\tau,\textbf{k}).
\end{equation}
We now show the properties of $F(\tau,\textbf{k})$ under
$P$ and $T$ transformations.
For $S=0$ from Eq.\ (2),
\begin{equation}
F(\textbf{k},\tau)
=  \epsilon_{\alpha \beta}[\theta_{\tau} \langle c_{\alpha, \textbf{k}}(\tau) c_{\beta, -\textbf{k}}(0)  \rangle - \theta_{-\tau} \langle  c_{\beta, -\textbf{k}}(0) c_{\alpha, \textbf{k}}(\tau)  \rangle  ],
\end{equation}
where $\theta_{\tau}$ is the Heaviside theta function.

We apply $PT$ to this $F$:
\begin{equation}\label{del-k-tau}
\begin{split}
& F(-\textbf{k},-\tau) \\
& =  \epsilon_{\mu \nu}[\theta_{-\tau} \langle c_{\mu, -\textbf{k}}(-\tau) c_{\nu, \textbf{k}}(0)  \rangle - \theta_{\tau} \langle  c_{\nu, \textbf{k}}(0) c_{\mu, -\textbf{k}}(-\tau) \rangle]  \\
& =  \epsilon_{\mu \nu}[\theta_{-\tau} \langle c_{\mu, -\textbf{k}}(0) c_{\nu, \textbf{k}}(\tau)  \rangle - \theta_{\tau} \langle  c_{\nu, \textbf{k}}(\tau) c_{\mu, -\textbf{k}}(0) \rangle ],
\end{split}
\end{equation}
where in the last line we have used the fact that $\langle T A(-\tau) B(0) \rangle = \langle T A(0) B(\tau) \rangle$ which agrees with the cyclicity of the trace,
\begin {eqnarray}
\langle A(-\tau)B(0) \rangle &=& {\rm Tr}( e ^{-H\tau} A e ^{H\tau} B ) = \nonumber\\
 {\rm Tr}( A e ^{H\tau} B e ^{-H\tau} ) &=& \langle  A(0) B(\tau) \rangle.
\end{eqnarray}
Going back to Eq.\ \ref{del-k-tau}, we permute $\mu \leftrightarrow  \nu$,
\begin{equation}\label{del'ktau}
\begin{split}
& F(-\textbf{k},-\tau) \\
& =  \epsilon_{\mu \nu}[\Theta_{\tau} \langle c_{\mu, \textbf{k}}(\tau) c_{\nu, -\textbf{k}}(0)  \rangle -
\Theta_{-\tau} \langle  c_{\nu, -\textbf{k}}(0) c_{\mu, \textbf{k}}(\tau)  \rangle  ] \\
& = F( \textbf{k},\tau).
\end{split}
\end{equation}
All these properties of the Gor'kov function will be reflected in the behavior of the gap function as well. Therefore, the gap function in general is even only under simultaneous
transformation: $\textbf{k} \to -\textbf{k}$ ($P$) and
$\tau \to -\tau $ ($T$). We recall that $PT=1$
 is not only satisfied by $P=+1$, $T=+1$
but also by $P=-1$ and $T=-1$. The former describes
the BCS $s$-wave (even-frequency) pairing whereas the
latter describes odd-frequency pairing.

\section{hamiltonian and wave function}
When the idea of the odd-frequency pairing was first formulated for
 the singlet superconductor, an effective spin-independent interaction
  mediated by phonon was considered. \cite{abrahams1993} It was
realized that this kind of interaction was unphysical for the singlet
 pairing.\cite{abrahams1993} The problem was solved by considering
 spin dependent electron-electron interactions. Odd-frequency pairing
posed another problem related to the selection of the order parameter. In
the BCS case the order parameter is generated from the expectation value,
 $F(\textbf{r},t;\textbf{r}',t' \to t)= \langle \psi(\textbf{r},t)\psi(\textbf{r}',t) \rangle$.
 But for the odd-frequency superconductor the equal-time gap vanishes since the gap is
 odd in frequency. This problem was solved by taking
$d F(\textbf{r},t;\textbf{r}',t')/dt|_{t \to t'}$ as the equal-time order
 parameter. \cite{abrahams1995}

A Hamiltonian having a spin dependent electron-electron interaction
was introduced by Abrahams  et al.  \cite{abrahams1995}. Using the
equation of motion they derived an expression for $d
F(\textbf{r},t;\textbf{r}',t')/dt|_{t \to t'}$. It was shown that
the equal-time condensate for odd-frequency pairing is the
expectation value of the product of a pair operator and a spin
excitation operator. In what follows, we adopt this approach, but
for an odd frequency $s$-wave $m=1$ triplet phase. We rewrite the
Hamiltonian from Ref. 2 in the following form, \beq
H&=&\sum_{\textbf{k}}\epsilon_{\textbf{k}\uparrow }c_{\textbf{k}\uparrow}^{\dag}c_{\textbf{k}\uparrow}+\sum_{\textbf{k}}\epsilon_{\textbf{k}\downarrow }c_{\textbf{k}\downarrow}^{\dag}c_{\textbf{k}\downarrow}+\sum_{\textbf{q}}\omega_{\textbf{q}}S_{\textbf{q}}^{+}S_{\textbf{q}}^{-} \nonumber \\
&+& \sum_{\textbf{k}\textbf{l}\textbf{q}\textbf{p}} V_{\textbf{k}\textbf{l}\textbf{q}\textbf{p}}c_{\textbf{k}+\frac{\textbf{q}}{2}\uparrow}^{\dag}c_{-\textbf{k}+\frac{\textbf{q}}{2}\downarrow}^{\dag}S_{\textbf{q}}^{+}c_{-\textbf{l}+\frac{\textbf{p}}{2}\downarrow}c_{\textbf{l}+\frac{\textbf{p}}{2}\uparrow}S_{p}^{-},
\label{hamiltonian}
\eeq
where $\epsilon_{\textbf{k}\uparrow \downarrow}$ refers to the kinetic energy of the $\uparrow \downarrow$ electrons measured from the Fermi energy, $\omega_\textbf{q}$ is the magnon kinetic energy, and $V_{\textbf{k}\textbf{l},\textbf{q}\textbf{p}}$ is an attractive
interaction which mediates the condensation. $c_{\textbf{k}\sigma}^{\dag}$ and $c_{\textbf{k}\sigma}$ creates and annihilates electrons at the state $\textbf{k} \sigma$. $S^{\pm}$ describe magnon excitations. Using this Hamiltonian, we propose a BCS-like wave function and study the superconducting state.

The proposed wave function is written as
\begin{equation}
| \psi  \rangle = \prod_{\textbf{k}\textbf{q}}
(u_{\textbf{k}\textbf{q}}+ v_{\textbf{k}\textbf{q}}
c_{\textbf{k}+\frac{\textbf{q}}{2}\uparrow}^{\dag}
c_{-\textbf{k}+\frac{\textbf{q}}{2}\downarrow}^{\dag}S_{\textbf{q}}^{+})
| 0 \rangle, \label{wavefunction}
\end{equation}
where $ | 0 \rangle$ represents the vacuum for both the electrons
and the spin bosons. This wave function describes the superposition
of the wave functions having two paired electrons with
$\textbf{k}+\frac{\textbf{q}}{2}$ and
$-\textbf{k}+\frac{\textbf{q}}{2}$ momentum and carrying opposite
spins and condensed along with spin excitations ($S_\textbf{q}^+$).
$v_{\textbf{k}\textbf{q}}$ ($u_{\textbf{k}\textbf{q}}$) represent
the amplitude of the occupation (or unoccupation) of these electron
pairs with the spin excitation.

There are key properties that explain this particular choice of
variational function: i) $| \psi  \rangle$ is a coherent state of
composite bosons ($
c_{\textbf{k}+\frac{\textbf{q}}{2}\uparrow}^{\dag}
c_{-\textbf{k}+\frac{\textbf{q}}{2}\downarrow}^{\dag}S_{\textbf{q}}^{+})$)
that carry charge $2e$; ii) this wave function describes  a coherent
state that has broken $U(1)$ symmetry associated with
superconducting condensate, as can be explicitly verified by using
$c_{\textbf{k}} \rightarrow \exp(i\phi)c_{\textbf{k}}$; iii)
Composite boson that condenses is {\em not a simple Cooper pair}
\cite{abrahams1995} but contains  two fermions and a spin-1 boson;
iv) composite boson field has finite expectation value in this state
\beqa
 \langle \psi |
c_{\textbf{k}+\frac{\textbf{q}}{2}\uparrow}^{\dag}
c_{-\textbf{k}+\frac{\textbf{q}}{2}\downarrow}^{\dag}S_{\textbf{q}}^{+}
| \psi  \rangle = u_{\textbf{k}\textbf{q}}v_{\textbf{k}\textbf{q}}
\label{EQ:coherentorder1} \eeqa
 and therefore $| \psi  \rangle$ is a mean field wave function for the composite
 condensate.

 The normalization of the wave function is given by,
\begin{equation}
\langle \psi | \psi \rangle = \prod_{\textbf{k}\textbf{q}} (|u_{\textbf{k}\textbf{q}}|^2+|v_{\textbf{k}\textbf{q}}|^2 \langle S^-S^+ \rangle _\textbf{q})=1,
\end{equation}
which implies that $|u_{\textbf{k}\textbf{q}}|^2+|v_{\textbf{k}\textbf{q}}|^2 \langle S^-S^+ \rangle _\textbf{q}=1$ for all $\textbf{k},\textbf{q}$.

To make a next step we need to find the expectation value of the
Hamiltonian (Eq.\ \ref{hamiltonian}) with respect to the wave
function (Eq.\ \ref{wavefunction}) and minimize it. Then we will
proceed to derive the quasi-particle dispersion, density of states,
and the self-consistent equation for the order parameter.

\section{total energy and its minimization}
The calculation of each term in Eq. \ref{hamiltonian} is shown in Appendix.
Using Eqs. \ref{UPKE}, \ref{DNKE}, \ref{MAGKE}, \ref{INTEN}, the total energy can be written as
\beq
E &= &\sum_{\textbf{k}\textbf{q}}( \epsilon_{\textbf{k}+\frac{\textbf{q}}{2}}+\epsilon_{\textbf{k}-\frac{\textbf{q}}{2}}+   \omega_{\textbf{q}} \langle S^-S^+ \rangle _\textbf{q} ) |v_{\textbf{k}\textbf{q}}|^2 \langle S^-S^+ \rangle _\textbf{q} \nonumber \\
&+& \sum_{\textbf{k}\textbf{l}\textbf{q}\textbf{p}}V_{\textbf{k}\textbf{l}\textbf{q}\textbf{p}}v_{\textbf{k}\textbf{q}}^*u_{\textbf{k}\textbf{q}}v_{\textbf{l}\textbf{p}}u_{\textbf{l}\textbf{p}}^* {\langle S^-S^+ \rangle_\textbf{q}} {\langle S^- S^+ \rangle_\textbf{p}}.
\eeq
Following the BCS method, we choose $u_{\textbf{k}\textbf{q}}, v_{\textbf{k}\textbf{q}}$ such that
they satisfy the normalization condition so that
$u_{\textbf{k}\textbf{q}}=\sin\theta_{\textbf{k}\textbf{q}}$ and $v_{\textbf{k}\textbf{q}}=\cos\theta_{\textbf{k}\textbf{q}}/\sqrt{
\langle S^-S^+ \rangle _\textbf{q}}$. Then the expression for the energy
reads
\begin{equation}
\begin{split}
E=\sum_{\textbf{k}\textbf{q}}  \cos^2\theta_{\textbf{k}\textbf{q}}  (\epsilon_{\textbf{k}+\frac{\textbf{q}}{2}}  + \epsilon_{\textbf{k}-\frac{\textbf{q}}{2}}+ \omega_{\textbf{q}} \langle S^-S^+ \rangle _\textbf{q}) +
 \\   \frac{1}{4}\sum_{\textbf{k}\textbf{l}\textbf{q}\textbf{p}}V_{\textbf{k}\textbf{l}\textbf{q}\textbf{p}} \sin 2\theta_{\textbf{k}\textbf{q}} \sin 2\theta_{\textbf{l}\textbf{p}} \sqrt{\langle S^- S^+ \rangle_\textbf{q}} \sqrt{\langle S^- S^+ \rangle_\textbf{p}}.
\end{split}
\end{equation}
The minimization of the energy with respect to $\theta_{\textbf{k}\textbf{q}}$ gives
\begin{equation}
\begin{split}
\frac{\partial E}{\partial\theta_{\textbf{k}\textbf{q}}}= -\sin 2\theta_{\textbf{k}\textbf{q}}
(\epsilon_{\textbf{k}+\frac{\textbf{q}}{2}}  + \epsilon_{\textbf{k}-\frac{\textbf{q}}{2}}+ \omega_{\textbf{q}}
\langle S^-S^+ \rangle _\textbf{q})  +
   \\  \sum_{\textbf{l}\textbf{p}}V_{\textbf{k}\textbf{l}\textbf{q}\textbf{p}} \cos 2\theta_{\textbf{k}\textbf{q}} \sin 2\theta_{\textbf{l}\textbf{p}}  \sqrt{\langle S^- S^+ \rangle_\textbf{q}} \sqrt{\langle S^- S^+ \rangle_\textbf{p}} =0,
\end{split}
\end{equation}
which can be rewritten as
\begin{equation}
\tan 2\theta_{\textbf{k}\textbf{q}} = \frac{ \sum_{\textbf{l}\textbf{p}}V_{\textbf{k}\textbf{l}\textbf{q}p}  \sin 2\theta_{\textbf{l}p} \sqrt{\langle S^- S^+ \rangle_\textbf{q}} \sqrt{\langle S^- S^+ \rangle_\textbf{p}}}{\epsilon_{\textbf{k}+\frac{\textbf{q}}{2}}  + \epsilon_{\textbf{k}-\frac{\textbf{q}}{2}}+ \omega_{\textbf{q}} \langle S^-S^+ \rangle _\textbf{q}}.
\end{equation}
We proceed by defining
 the two quantities  $\Delta$
 and $E$ that will turn out to
  be the gap parameter and the energy of a composite excitation.
\begin{subequations}
\begin{equation}
\Delta_{\textbf{k}\textbf{q}}=-\frac{1}{2}\sum_{\textbf{l}\textbf{p}}V_{\textbf{k}\textbf{l}\textbf{q}\textbf{p}}  \sin 2\theta_{\textbf{l}\textbf{p}}  \sqrt{\langle S^- S^+ \rangle_\textbf{q}} \sqrt{\langle S^- S^+ \rangle_\textbf{p}} ,
\label{deltakq}
\end{equation}
\beq
E_{\textbf{k}\textbf{q}}&=&\sqrt{ (\frac{\epsilon_{\textbf{k}+\frac{\textbf{q}}{2}}  + \epsilon_{\textbf{k}-\frac{\textbf{q}}{2}}}{2}+ \frac{\omega_{\textbf{q}}}{2} \langle S^-S^+ \rangle _\textbf{q})^2+\Delta_{\textbf{k}\textbf{q}}^2} \nonumber \\
&=&\sqrt{(\epsilon_\textbf{k}+\frac{q^2}{8m}+\frac{\omega_{\textbf{q}}}{2} \langle S^-S^+ \rangle _\textbf{q})^2+\Delta_{\textbf{k}\textbf{q}}^2}.
\eeq
\end{subequations}
Then
\begin{subequations}
\begin{equation}
\sin 2\theta_{\textbf{k}\textbf{q}}=2 u_{\textbf{k}\textbf{q}} v_{\textbf{k}\textbf{q}}\sqrt {\langle S^- S^+ \rangle_\textbf{q}}=
\frac{\Delta_{\textbf{k}\textbf{q}}}{E_{\textbf{k}\textbf{q}}},
\label{sin2theta}
\end{equation}
and
\begin{equation}
\cos 2\theta_{\textbf{k}\textbf{q}}= v_{\textbf{k}\textbf{q}}^2 \langle S^- S^+ \rangle _\textbf{q} -
u_{\textbf{\textbf{k}}\textbf{q}}^2=\frac{- \xi_{\textbf{k}\textbf{q}}}{E_{\textbf{k}\textbf{q}}},
\end{equation}
\label{probabilityunoccupied}
\end{subequations}
where we have introduced the abbreviation
\begin{equation}
\xi_{k\textbf{q}}= \frac{\epsilon_{\textbf{k}+\frac{\textbf{q}}{2}}  + \epsilon_{\textbf{k}-\frac{\textbf{q}}{2}}}{2}+ \frac{\omega_{\textbf{q}}}{2} \langle S^-S^+ \rangle _\textbf{q}.
\end{equation}

Solving the normalization condition and Eq.\
\ref{probabilityunoccupied}, we can show that,
\begin{subequations}
\begin{equation}
u_{\textbf{k}\textbf{q}}^2=\frac{1}{2}(1+\frac{\xi_{\textbf{k}\textbf{q}}}{E_{\textbf{k}\textbf{q}}}),
\end{equation}
\begin{equation}
v_{\textbf{k}\textbf{q}}^2=\frac{1}{2 \langle S^- S^+ \rangle _\textbf{q}}(1-\frac{\xi_{\textbf{k}\textbf{q}}}{E_{\textbf{k}\textbf{q}}})
\label{probability}
\end{equation}
\end{subequations}

BCS limit can be recovered at any stage of this analysis if we
assume that spin correlators are factorized and have a peak at ${\bf
q} = 0$. This limit corresponds to the condensation of spin field $
\langle S^- S^+ \rangle _\textbf{q} = \langle S^-  \rangle
_\textbf{q} \langle   S^+ \rangle _\textbf{q} \delta_{{\bf q}, 0}$.
In this limit additional summation over $\bf q$ drops out and we
recover standard BCS logarithm in selfconsistency equation
Eq.(\ref{deltakq}), along with other features of BCS solution. This
limit   corresponds to the factorizitation of composite boson into
product $ \langle \psi |
c_{\textbf{k}+\frac{\textbf{q}}{2}\uparrow}^{\dag}
c_{-\textbf{k}+\frac{\textbf{q}}{2}\downarrow}^{\dag}S_{\textbf{q}}^{+}
| \psi \rangle \rightarrow \langle \psi | c_{{\textbf{k}}
\uparrow}^{\dag} c_{{-\textbf{k}}\downarrow}^{\dag} | \psi \rangle
\langle \psi  |S_{\textbf{q}}^{+} | \psi \rangle \delta_{\bf q,0}$.

\section{energy spectrum}

Unlike the BCS case, $E_{\textbf{k}\textbf{q}}$ is not a single-particle excitation energy. Therefore, we shall derive an expression for the energy required to excite an
electron from the superconducting ground state. The excited state
for an up spin  is given by,
\begin{equation}
\widetilde{\psi}_{\uparrow}=[\prod_{\textbf{q},\textbf{k} \neq \textbf{k}'}(u_{\textbf{k}\textbf{q}}+v_{\textbf{k}\textbf{q}}
b_{\textbf{k}\textbf{q}}^{\dag})] c_{\textbf{k}'+\frac{\textbf{q}}{2}\uparrow}^{\dag}  | 0 \rangle,
\label{excite_wf}
\end{equation}
where we have defined the composite creation operator $b_{\textbf{k}\textbf{q}}^{\dag}=c_{\textbf{k}+\frac{\textbf{q}}{2}\uparrow}^{\dag} c_{-\textbf{k}+\frac{\textbf{q}}{2}\downarrow}^{\dag}S_{\textbf{q}}^{+}$. We  calculate the expectation value of the Hamiltonian Eq.\ (\ref{hamiltonian}) with respect to the excited state wave function Eq.\ (\ref{excite_wf}). The details are in Appendix E.
The expectation value can be expressed as\begin{equation}
\langle \widetilde{\psi}_{\uparrow} |H|\widetilde{\psi}_{\uparrow} \rangle =
\langle \psi |H| \psi \rangle + \epsilon_{\textbf{k}'+\frac{\textbf{q}}{2}}+\frac{\Delta_{\textbf{k}'\textbf{q}}^2}{E_{\textbf{k}'\textbf{q}}}- 2\xi_{\textbf{k}'\textbf{q}} v_{\textbf{k}'\textbf{q}}^2 \langle S^- S^+ \rangle_\textbf{q}.
\end{equation}
Using Eq.\ (\ref{probability}), we can rewrite the above equation as,
\begin{equation}
 \Delta E_{\uparrow}= \epsilon_{\textbf{k}'+\frac{\textbf{q}}{2}} -\xi_{\textbf{k}'\textbf{q}}+E_{\textbf{k}'\textbf{q}},
 \label{qp1}
\end{equation}
where $ \Delta E_{\uparrow}=\langle \widetilde{\psi}_{\uparrow}
|H|\widetilde{\psi}_{\uparrow}\rangle-\langle \psi |H| \psi
\rangle$ is the excitation energy of the up spin electrons.  $\Delta E_{\uparrow}$ can also be written as
$\Delta E_{\uparrow}=\textbf{k}' \cdot
\textbf{q}/2m^*-(\omega_{\textbf{q}}/2) \langle S^- S^+
\rangle_\textbf{q}+E_{\textbf{k}'\textbf{q}}$.
Doing the same for the down spin excited state $\widetilde{\psi}_{\downarrow}$, we find
$\Delta E_{\downarrow}= - \textbf{k}' \cdot
\textbf{q}/2m^*-(\omega_{\textbf{q}}/2) \langle S^- S^+
\rangle_\textbf{q}+E_{\textbf{k}'\textbf{q}}$.

\section{density of states}

\begin{figure*}
\includegraphics[width=9.0cm, height=14.0cm,angle=-90]{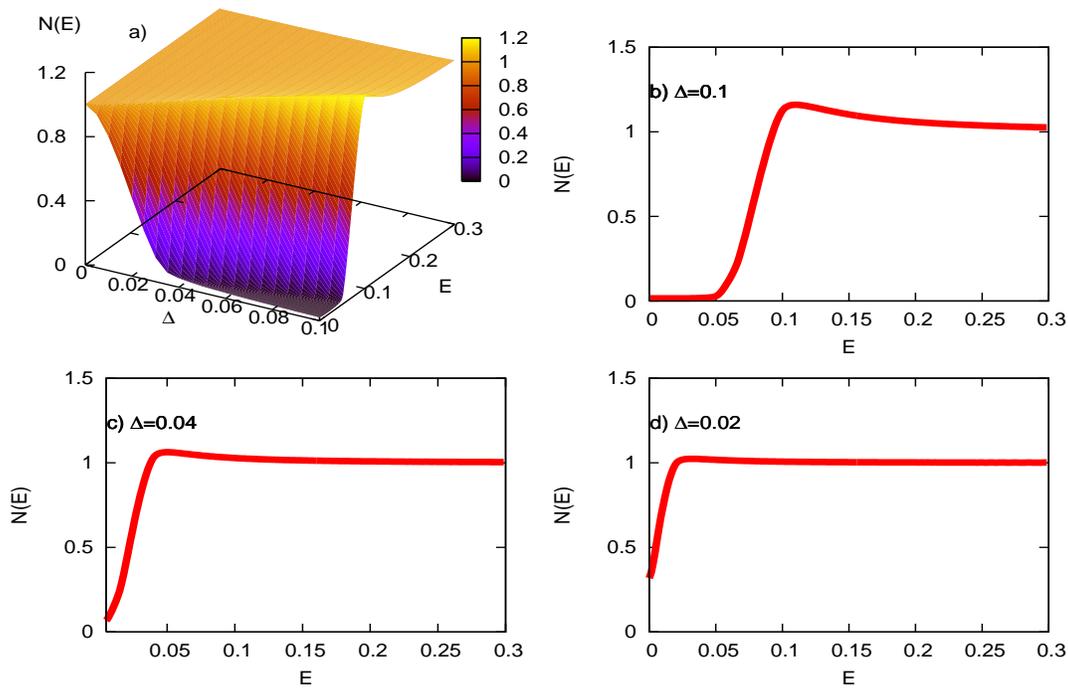}
\caption{The density of states (DOS) as a function of
energy and superconducting order parameter in an odd-frequency
superconductor. The DOS is normalized with respect to the DOS of the
normal state. All the energies are normalized with Fermi
energy of the system.  The result in the upper panel is presented
for the magnon momentum cutoff $q_c=0.25$. In this figure we show
that the density of state is finite even for energy less than the
gap energy. The maximum of the DOS is at the gap edge but the DOS is
highly reduced compared to the BCS case. For smaller gap energy the
DOS is completely gapless. Once the gap is closed the DOS starts to
pile up at $E=0$ for smaller values of $\Delta$.} \label{dos}
\end{figure*}

The density of states (DOS) as a function of energy, $N(E)$, is defined as,
\begin{equation}
\label{dos_numerically}
\begin{split}
N_{\pm}(E)=  \sum_{\textbf{k}\textbf{q}}  \delta[E-(\pm \frac{\textbf{k} \cdot \textbf{q}}{2m^*}-\frac{\omega_\textbf{q}}{2} \langle S^- S^+ \rangle_\textbf{q} +E_{\textbf{k}\textbf{q}})] ,
\end{split}
\end{equation}
where $\pm$ corresponds to up and down spins respectively. We numerically calculate the density of states for two cases of the magnon dispersion: 1) $\omega_\textbf{q}=\textbf{q}^2/2M$, and 2)
$\omega_\textbf{q}=\omega_0$.. We set  $ \langle S^-S^+ \rangle _\textbf{q}=1.0$, and $M=10m^*$. The DOS for
case 1) is shown in Fig.\ \ref{dos}.

In Fig \ref{dos}a we show the DOS as a function of energy and order parameter. We have set
a magnon momentum cutoff, $q_c=0.25k_F$.
We see that the DOS can be non-zero for energies less than the superconducting gap parameter; hence the DOS is gapless.  The maximum of the density of state is always at the gap edge, but it is highly reduced at the gap edge compared to the BCS
case. At $E=0$, the DOS can be non-zero for small $\Delta$. The calculation for a smaller $q_c$ (not shown in the figure) shows that the gap becomes more prominent in the DOS and spectral weight is transferred to the gap edge, similar to the BCS case. Hence $q_c \to 0$ reproduces the BCS results. In Fig.\ \ref{dos}b-d we have shown the plane cut of Fig.\ \ref{dos}a for different values of $\Delta$. For $\Delta=0.1$ (Fig.\ \ref{dos}b) we see that the DOS is non-zero for $0.05<E<\Delta$. For $\Delta=0.04$ (Fig.\ \ref{dos}c) we see that the gap is completely closed and the excitations will be gapless. The effect is even bigger for $\Delta=0.02$.

\begin{figure}[h]
\includegraphics[width=5.0cm, height=8.0cm,angle=-90]{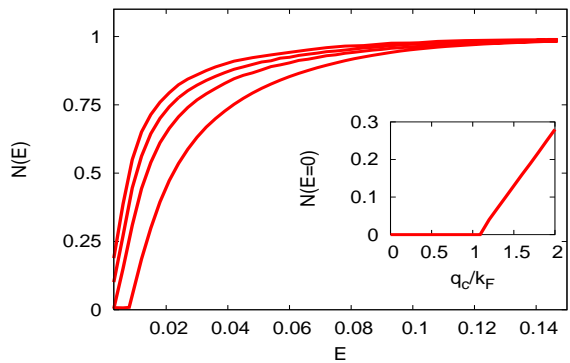}
\caption{ The DOS at fixed $\Delta=0.1$ as a function of energy for  the magnon momentum cutoff $q_c \ge k_F$. The values of $q_c$ are $0.9,1.1,1.4,1.7$ times
$k_F$. As we increase $q_c$ the gap in the density of states gradually closes up. For bigger $q_c$ the DOS piles up at $E=0$.  }
\label{dos_bigqc}
\end{figure}

We also calculated the DOS using case 2): $\omega_\textbf{q}=\textbf{q}^2/2M$ for $q_c \ge k_F$ (the Fermi momentum) for a fixed value of $\Delta=0.1$. The result is shown in Fig.\ \ref{dos_bigqc}. In this figure we can see that the DOS almost closes the gap when  $q_c=k_F$. As we increase $q_c$, the gap closes completely. Then the quasiparticle excitations become gapless. A still further increase in $q_c$ results in a finite DOS at $E=0$. For $q_c \ge k_F$ there is  no enhancement of the spectral weight at the gap edge.

The calculation of the DOS for  $\omega_\textbf{q}=\omega_0$ also shows the similar density of state as discussed above for
both $q_c=0.25k_F$ and $q_c \ge k_F$.

\section{superconducting gap vs coupling constant}

The self-consistent gap equation (Eq.\ \ref{deltakq}) can be written as,
\begin{equation}\label{geq}
\begin{split}
\Delta_{\textbf{k}\textbf{q}}
=\frac{V}{2} \sum_{\textbf{l}\textbf{p}}\frac{\Delta_{\textbf{l}\textbf{p}}}{E_{\textbf{l}\textbf{p}}}\sqrt{\langle S^- S^+ \rangle_\textbf{q}} \sqrt{\langle S^- S^+ \rangle_\textbf{p}},
\end{split}
\end{equation}
where we have taken
\begin{equation}
V_{\textbf{k}\textbf{l}\textbf{p}\textbf{q}}= \Big \{ \begin{array}{r@{\quad : \quad} l} -V & | \epsilon_\textbf{k} | \leq 0.2 \mu  \\ 0 & | \epsilon_\textbf{k}| > 0.2\mu
\end{array}
\end{equation}
Then $\Delta_{\textbf{k}\textbf{q}}=\Delta=\Delta_{\textbf{l}\textbf{p}}$. The use of a more complicated interaction potential with a momentum dependence.
would bring additional calculational complications, which would not change the nature of the results.

We denote $p^2/8m^*+(\omega_{\textbf{p}}/2) \langle S^-S^+ \rangle _\textbf{p}$
by $f(p)$. We first perform the energy integral in Eq.\ \ref{geq} as follows,
\beq\label{gapeqn1}
1&=&\frac{V}{2} \sum_{lp}\frac{\langle S^- S^+ \rangle_\textbf{p}}{\sqrt{(\epsilon_\textbf{p}+f(p))^2+\Delta^2  }} \nonumber \\
&=&g \int  p^2 dp \int_0^{\hbar \omega_c}
\frac{\langle S^- S^+ \rangle_\textbf{p} d\epsilon }{\sqrt{(\epsilon_\textbf{p}+f(p))^2+\Delta^2 }} \nonumber  \\
&=&g \int   p^2 dp \langle S^- S^+ \rangle_\textbf{p}
\log\frac{ \epsilon_c(p)  + \sqrt{\Delta^2 + \epsilon_c(p)^2} }
{ f(p)+\sqrt{\Delta^2+f(p)^2}},\nonumber \\
\eeq
where $N(0)$ is the DOS in the  normal state at the
 Fermi energy, $g$ is the dimensionless coupling
  $N(0)V/2\pi^2$, $\epsilon_c(p)=\hbar \omega_c +  f(p)$ and $\hbar \omega_c
  =0.2\mu$. If we assume spin correlator to have a sharp peak
  $\delta_{{\bf q},0}$ we recover BCS selfconsistency equation from this equation.

In the BCS case the gap equation is $1=N(0)V \log[(\hbar \omega_c
+\sqrt{\Delta^2+\hbar \omega_c^2})/\Delta]$. There is a solution for
$\Delta$ for an arbitrary small   value of $N(0)V$ due to
logarithmic divergence of the integral. In our case, in the presence
of the magnon, the denominator will have some nonzero value
 because of the non-zero magnon energy.
 Then the right hand side can be made equal to 1 only for some
 critical value of $g$, as can be seen in the numerical evaluation discussed below.

\subsection{Case 1), $\omega_\textbf{p}=p^2/2M$}

\begin{figure}[h]
\includegraphics[width=5.0cm, height=8.0cm,angle=-90]{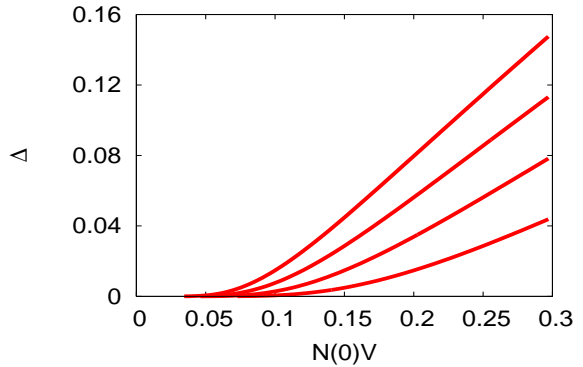}
\caption{The order parameter is numerically calculated for $\omega_q=q^2/2M$. The magnon momentum cutoff is
given by $q_c=B k_F$, where
 $B=0.12, 0.1,0.08,0.06$, top to bottom.
The order parameter is non zero only for critical value of the
coupling $g=N(0)V/2\pi^2$.  For a given value of $N(0)V$ the larger $\Delta$ corresponds to
the larger magnon momentum cutoff. }
\label{good_phasediagram_variableqc}
\end{figure}

We solve Eq.\ \ref{gapeqn1} numerically for $\Delta$
 as a function of the coupling strength
$g$. We set $m^*/M=0.1$ and $\langle S^-S^+  \rangle_\textbf{p}=1.0$. The cutoff for the
magnon momentum is given by $q_c=B k_F$, where $B$ varies between
$0.12$ to $0.06$ in equal steps of $0.02$. The result is shown
in Fig.\ \ref{good_phasediagram_variableqc}. In this figure we can
see that a nonzero order parameter requires a critical coupling.

\subsection{Case 2), $\omega_{\textbf{p}}=\omega_0$}

The gap equation is again given by Eq.\ \ref{gapeqn1} but now
\begin{equation}
f(p) = p^2/8m^* -  \omega_0 \langle S^-S^+  \rangle_\textbf{p}/2.
\end{equation}

\begin{figure}[h]
\includegraphics[width=5cm, height=8.0cm,angle=-90]{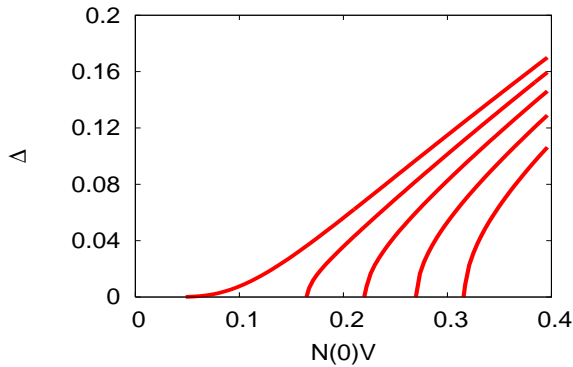}
\caption{The order parameter is calculated for $\omega_q=\omega_0$. The
 magnon momentum cutoff is $q_c=0.1k_F$. The result is shown for $\omega_0=C \mu$ where
$C=0,0.02,0.04,0.06,0.8$, top to bottom.
The order parameter is non-zero only for coupling exceeding a critical value.
 For a given value of $N(0)V$ the bigger $\Delta$ corresponds to the smaller value of $\omega_0$. }
\label{good_phasediagram_variableomega}
\end{figure}

We solve Eq.\ \ref{gapeqn1} numerically for $\Delta$ as a function of the coupling strength $g$. We fix the cutoff for
the magnon momentum to be $0.1k_F$. The result
for various $\omega_0 =C \mu$ where $C=0,0.02,0.04,0.06,0.08$ is shown in
Fig. \ref{good_phasediagram_variableomega}.  Again, the superconducting transition requires a critical
coupling.

\section{Meissner effect}

The Meissner effect is one of the defining properties of a superconductor. The Meissner effect has been derived for the composite odd-frequency superconductor by Abrahams et al.\ \cite{abrahams1995}. Here, we summarize the derivation given in that reference

A superconductor shows the Meissner effect when the paramagnetic electrodynamic response is less than the diamagnetic response. The dc response is given by,
\begin{equation}
\begin{split}
j_i(q)=-Q_{ij}(\textbf{q})A_j(\textbf{q}), \\
Q_{ij}(\textbf{q})=\delta_{ij}\frac{Ne^2}{m}+Q^p_{ij}(\textbf{q}),
\end{split}
\end{equation}
where $A(q)$ is the Fourier transform of vector potential $A(r)$, $N$ is the electron density, and $m$ is their mass. $Q^p_{ij}(\textbf{q})$ is given by,
\begin{widetext}
\begin{equation}
Q^p_{ij}(\textbf{q})= \frac{-e^2}{4m^2}\sum_{\gamma \delta}  \sum_{kk'} \textbf{k}_i \textbf{k}'_j
\int_{-\beta}^{\beta} d\tau  \langle Tc^{\dag}_{\gamma}(\textbf{k}_+,\tau)c_{\gamma}(\textbf{k}_-,\tau)c^{\dag}_{\delta}(\textbf{k}'_-,0)c_{\delta}(\textbf{k}'_+,0) \rangle ,
\end{equation}
\end{widetext}
where $\textbf{k}_{\pm}=\textbf{k} \pm \textbf{q}/2$. $Q^p$ can be evaluated near the critical temperature $T_c$ by perturbation in the order parameter $\Delta$. The relevant Feynman diagrams of the current-current correlation function for the Meissner effect are used. The analytical expression for $q \to 0$ is
\begin{widetext}
\begin{equation}
Q^p_{ij}(\textbf{q})-Q^n_{ij}(\textbf{q})= \frac{e^2T^2\Delta^2}{m^2}\sum_{\omega \omega', \textbf{k}\textbf{k}'} \textbf{k}_i \textbf{k}'_j
 [  G^2(\textbf{k},\omega)G^2(\textbf{k}',\omega')-2G^3(\textbf{k},\omega)G(\textbf{k}',\omega')  ] D(\textbf{k}+\textbf{k}',\omega+\omega'),
\end{equation}
\end{widetext}
where, $G(\textbf{k},\omega)$ and $D(\textbf{k},\omega)$ are the electron and magnon propagators. The condition for the Meissner effect is given by $Q^p - Q^n > 0$, which signifies the positive superfluid density in the superconductor.

Situations with several models of the magnon propagators are discussed. If the magnon propagator is momentum independent, there is no contribution to $Q_{ij}(\textbf{q})$ since the momentum summands are odd functions. So a momentum-dependent magnon propagator is used to discuss the Meissner effect. In the case of a static, spatially uniform magnon propagator having factorized form given by, $D(q,\nu)=-\delta_\textbf{q} \delta_\nu$, the Meissner effect is found ($Q^p - Q^n > 0$).  For spread-out $\delta$-functions, the sign of $Q^p - Q^n $ does not change, thus a positive superfluid density with a value between zero and the BCS value. Thus, it is shown that the composite odd-frequency superconductors exhibit the Meissner effect.

\section{Conclusion}

In this paper we propose a BCS-like wave function for the s-wave
triplet odd frequency superconductor. Our alternative approach to
the odd-frequency superconductivity  is based on the earlier
discussion on composite bosons \cite{abrahams1995}. We present the
wave function for the odd frequency superconductor $| \psi  \rangle
= \prod_{\textbf{k}\textbf{q}} (u_{\textbf{k}\textbf{q}}+
v_{\textbf{k}\textbf{q}}
c_{\textbf{k}+\frac{\textbf{q}}{2}\uparrow}^{\dag}
c_{-\textbf{k}+\frac{\textbf{q}}{2}\downarrow}^{\dag}S_{\textbf{q}}^{+})
| 0 \rangle, $ , Eq.(\ref{wavefunction}), that explicitly contains
only the equal time operators and hence does not involve frequency
or time domain. The wave function describes a condensate of a Cooper
pair of spin $S=0$ and a magnon of spin $S=1$.   $|\psi \rangle$
does describe a coherent state that has nonzero expectation value
for the composite boson operator, it captures  the charge $2e$
condensate that breaks gauge symmetry and corresponds to the
superconducting state. Naturally, since this $|\psi \rangle$
describes odd-frequency superconductor, spatial parity P of this
condensate is reversed compared to the even frequency pairing
operators that corresponds to BCS condensate.  Specifically, for the
case we considered of spin triple $S=1$ odd frequency condensate the
spatial parity of the composite boson $\langle
c^{\dag}_{\uparrow}(\br) c^{\dag}_{\downarrow}(\br) S^+(\br)
\rangle$ is $P = +1$ and hence this order parameter does posess all
the quantum numbers inherent to the odd frequency $S = 1$
superconductor.

We present a simplified model that captures the important features
of the strong coupling theory developed for the odd-frequency
superconductors and our results agree with the predictions of
earlier studies: i) we show that the superconductivity requires a
critical coupling. It was argued earlier
\cite{berezinskii1974,balatsky1992} that a critical coupling is
necessary in order to get the superconducting transition in the odd
frequency superconductor, which we have also shown in this work. ii)
we also derive the dispersion relation for the quasiparticles. We
determine the density of states of the excitations. The density of
states is very different from that of the BCS case. The gapless
nature of  quasiparticle excitations we find  is also in agreement
with earlier predictions. The calculation of the density of states
shows that it is always higher at the gap edge but its magnitude is
highly reduced compared to the BCS case. For a range of parameters,
unlike the BCS case, the DOS is finite for energies less than the
gap energy and at $E=0$ it can be non-zero, hence odd-frequency
supercoductor is {\em gapless}. We also argues how the BCS result is
recovered by taking the magnon operator to condense and momentum
cutoff $q_c=0$.

Present discussion would be useful for the equal time formulation of
the odd-frequency superconducting state and physical observables
related to condensate. It also would be useful in elucidating the
nature of condensate in odd-frequency supercondutors.

Work at Los Alamos was supported by US DOE through LDRD and BES. We
also acknowledge hospitality of KITP at UC Santa Barbara.

\appendix

\section{Kinetic energy of up spin electrons}
It is convenient to rewrite $\sum_\textbf{k}  \epsilon_{\textbf{k}\uparrow}c_{\textbf{k}\uparrow}^{\dag}c_{\textbf{k}\uparrow} $ as,
\begin{equation}
( \frac{1}{N}\sum_q)\sum_\textbf{k}  \epsilon_{\textbf{k}\uparrow}c_{\textbf{k}\uparrow}^{\dag}c_{\textbf{k}\uparrow}= \frac{1}{N} \sum_{\textbf{k}\textbf{q}} \epsilon_{\textbf{k}+\frac{\textbf{q}}{2}\uparrow}c_{\textbf{k}+\frac{\textbf{q}}{2}\uparrow}^{\dag}c_{\textbf{k}+\frac{\textbf{q}}{2}\uparrow}.
 \end{equation}
This is a trivial identity since we can shift $\textbf{k} \to \textbf{k}+\frac{\textbf{q}}{2}$ and get the same result.

We denote the $(\textbf{m} \textbf{n})$ component of the wave function as, $|  \psi_{\textbf{m} \textbf{n}} \rangle=
(u_{\textbf{m} \textbf{n}}+v_{\textbf{m} \textbf{n}}c_{\textbf{m} +\frac{\textbf{n} }{2}\uparrow}^{\dag} c_{-\textbf{m} +\frac{\textbf{n} }{2}\downarrow}^{\dag}S_\textbf{n} ^{\dag})|0\rangle$.
The expectation value of the kinetic energy of the up spin electrons $KE_{\uparrow}=\frac{1}{N}\sum_{\textbf{k}\textbf{q}}
 \langle  \psi^*  |   \epsilon_{\textbf{k}+\frac{\textbf{q}}{2}\uparrow}c_{\textbf{k}+\frac{\textbf{q}}{2}\uparrow}^{\dag}c_{\textbf{k}+\frac{\textbf{q}}{2}\uparrow}  | \psi \rangle$ is given by,

\begin{equation}
\label{UPKE}
\begin{split}
KE_{\uparrow} =
\frac{1}{N}\sum_{\textbf{k}\textbf{q}}   \langle  \psi_{\textbf{k}'q'}^*  |   \epsilon_{\textbf{k}+\frac{\textbf{q}}{2}\uparrow}c_{\textbf{k}+\frac{\textbf{q}}{2}\uparrow}^{\dag}c_{\textbf{k}+\frac{\textbf{q}}{2}\uparrow}  | \psi_{\textbf{k}\textbf{q}} \rangle \\
= \frac{1}{N}\sum_{\textbf{k}\textbf{q}} \epsilon_{\textbf{k}+\frac{\textbf{q}}{2}} |v_{\textbf{k}\textbf{q}}|^2 \langle S^-S^+  \rangle _\textbf{q}.
 \end{split}
\end{equation}
Here we use the normalization condition that $ \langle \psi_{\textbf{m}' \neq \textbf{k}',\textbf{q} '}^* |  \psi_{\textbf{m}  \neq \textbf{k},\textbf{q}} \rangle  = \delta_{\textbf{m} \textbf{m} '} \delta_{\textbf{q}\textbf{q} '}$.

Then the kinetic energy of the up spin electrons is $KE_{\uparrow}=\sum_{\textbf{k}\textbf{q}} \epsilon_{\textbf{k}+\frac{\textbf{q}}{2}} |v_{\textbf{k}\textbf{q}}|^2 \langle S^-S^+  \rangle _\textbf{q} $.

\section{Kinetic energy of down spin electrons}
Using the same argument as discussed in appendix A, we rewrite,  $\sum_\textbf{k} \epsilon_{\textbf{k}\downarrow}c_{\textbf{k}\downarrow}^{\dag}c_{\textbf{k}\downarrow} $ as,
\begin{equation}
( \frac{1}{N}\sum_\textbf{q})\sum_\textbf{k}  \epsilon_{\textbf{k}\downarrow}c_{\textbf{k}\downarrow}^{\dag}c_{\textbf{k}\downarrow}= \frac{1}{N} \sum_{\textbf{k}\textbf{q}}  \epsilon_{-\textbf{k}+\frac{\textbf{q}}{2}\downarrow}c_{-\textbf{k}+\frac{\textbf{q}}{2}\downarrow}^{\dag}c_{\textbf{k}+\frac{\textbf{q}}{2}\downarrow}.
 \end{equation}

Then the expectation value of the kinetic energy of the down spin electrons $KE_{\downarrow}=\frac{1}{N}\sum_{\textbf{k}\textbf{q}}    \langle  \psi^*  |   \epsilon_{-\textbf{k}+\frac{\textbf{q}}{2}\downarrow}c_{-\textbf{k}+\frac{\textbf{q}}{2}\downarrow}^{\dag}c_{\textbf{k}+\frac{\textbf{q}}{2}\downarrow}  | \psi \rangle$ is given by

\begin{equation}
\label{DNKE}
\begin{split}
KE_{\downarrow} = \frac{1}{N}\sum_{\textbf{k}\textbf{q}}    \langle  \psi_{\textbf{k}'\textbf{q}'}^*  |   \epsilon_{-\textbf{k}+\frac{\textbf{q}}{2}\downarrow}c_{-\textbf{k}+\frac{\textbf{q}}{2}\downarrow}^{\dag}c_{-\textbf{k}+\frac{\textbf{q}}{2}\downarrow}  | \psi_{\textbf{k}\textbf{q}} \rangle \\
= \frac{1}{N}\sum_{\textbf{k}\textbf{q}} \epsilon_{-\textbf{k}+\frac{\textbf{q}}{2}} |v_{\textbf{k}\textbf{q}}|^2 \langle S^-S^+  \rangle _\textbf{q} .
\end{split}
\end{equation}

Then the kinetic energy of the down spin electrons is $KE_{\downarrow}=\frac{1}{N}\sum_{\textbf{k}\textbf{q}} \epsilon_{\textbf{k}-\frac{\textbf{q}}{2}} |v_{\textbf{k}\textbf{q}}|^2 \langle S^-S^+  \rangle _\textbf{q} $.

\section{Magnon energy}

The expectation value of the magnon kinetic energy $KE_m= \sum_\textbf{q} \langle \psi^* | \omega_\textbf{q} S^+_\textbf{q} S^-_\textbf{q} | \psi \rangle$  can be rewritten as
\begin{equation}
KE_m= (\frac{1}{N}\sum_\textbf{k}) \sum_\textbf{q}  \langle \psi_{\textbf{k}'q'}^* | \omega_\textbf{q} S^+_\textbf{q} S^-_\textbf{q} | \psi_{\textbf{k}\textbf{q}} \rangle,
\end{equation}
which gives,
\begin{equation}
\label{MAGKE}
\begin{split}
KE_m =
\frac{1}{N}\sum_{\textbf{k}\textbf{q}} \omega_{\textbf{q}} |v_{\textbf{k}\textbf{q}}|^2 \langle   S^- S^+  S^- S^+  \rangle_\textbf{q}
\end{split}
\end{equation}

\section{Interaction energy}
In the calculation of the expectation value of the interaction energy, $E_I$, it is easy to see that the product of only two states, $\textbf{k}\textbf{q}$ and $\textbf{l}\textbf{p}$ give non-zero contribution to the interaction term.
All the other states are normalized to unity.  Then,
\begin{equation}
\begin{split}
E_I= \frac{1}{N}\sum_{\textbf{k}\textbf{l}\textbf{p}\textbf{q}} \langle \psi^* |V_{\textbf{k}\textbf{l}\textbf{q}\textbf{p}} b_{\textbf{k}\textbf{q}}^{\dag}   b_{\textbf{l}\textbf{p}} | \psi \rangle \\
= \frac{1}{N} \sum_{\textbf{k}\textbf{l}\textbf{q}\textbf{p}} V_{\textbf{k}\textbf{l}\textbf{p}\textbf{q}}  \langle   \psi_{\textbf{k}\textbf{q}}^*  \psi_{\textbf{l}\textbf{p}}^*   | b_{\textbf{k}\textbf{q}}^{\dag}   b_{\textbf{l}\textbf{p}} |  \psi_{\textbf{l}\textbf{p}} \psi_{\textbf{k}\textbf{q}}   \rangle   \\
=   \frac{1}{N} \sum_{\textbf{k}\textbf{l}\textbf{q}\textbf{p}}  V_{\textbf{k}\textbf{l}\textbf{q}\textbf{p}} v_{\textbf{k}\textbf{q}}^* u_{\textbf{k}\textbf{q}} v_{\textbf{l}\textbf{p}} u_{\textbf{l}\textbf{p}}^* {\langle S^-S^+ \rangle_\textbf{q}} {\langle S^- S^+ \rangle_\textbf{p}}
\label{INTEN}
\end{split}
\end{equation}
where $b_{\textbf{k}\textbf{q}}^{\dag}=c_{\textbf{k}+\frac{\textbf{q}}{2}\uparrow}^{\dag}c_{-\textbf{k}+\frac{\textbf{q}}{2}\downarrow}^{\dag}S_{\textbf{q}}^{+}$

\section{Energy of excited states}

The wave function of an excited state is,
\begin{equation}
\widetilde{\psi}={\big [}\prod_{{\bf q},\textbf{k}\neq \textbf{k}' }(u_{\textbf{k}\textbf{q}}+v_{\textbf{k}\textbf{q}} b_{\textbf{k}\textbf{q}}^{\dag}){\big ]} c_{\textbf{k}'+\frac{\textbf{q}}{2}\uparrow}^{\dag}  | 0 \rangle.
\end{equation}
Using the procedure of the Appendix A, we calculate the kinetic energy of the up spin electrons ($\widetilde{KE}_{\uparrow}$) with respect to the excited state wave function:
\begin{equation}
\widetilde{KE}_{\uparrow}=\frac{1}{N}\sum_{{\bf q},\textbf{k}\neq \textbf{k}' } \epsilon_{\textbf{k}+\frac{\textbf{q}}{2}} |v_{\textbf{k}\textbf{q}}|^2 \langle S^-S^+  \rangle _\textbf{q}  + \epsilon_{\textbf{k}'+\frac{\textbf{q}}{2}}
\label{excited_UPKE}
\end{equation}
where the restriction on $\textbf{k}$ in the summation is inherited from the restriction imposed on the excited state wave function. $\epsilon_{\textbf{k}'+\frac{\textbf{q}}{2}}$ is due to the creation operator
$c_{\textbf{k}'+\frac{\textbf{q}}{2}\uparrow}^{\dag}$ which creates an up spin electron having unit probability of occupation in the state of momentum $\textbf{k}'+\frac{\textbf{q}}{2}$. We rewrite the Eq.\ \ref{excited_UPKE} in the following form,
\begin{equation}
\begin{split}
\widetilde{KE}_{\uparrow}=\frac{1}{N}\sum_{\textbf{k}\textbf{q}} \epsilon_{\textbf{k}+\frac{\textbf{q}}{2}} |v_{\textbf{k}\textbf{q}}|^2 \langle S^-S^+  \rangle _\textbf{q}   + \epsilon_{\textbf{k}'+\frac{\textbf{q}}{2}}
- \epsilon_{\textbf{k}'+\frac{\textbf{q}}{2}} |v_{\textbf{k}'\textbf{q}}|^2  .
\end{split}
\label{EUPKE}
\end{equation}

Proceeding similarly, we show that the kinetic energy of the down spin electrons can be written as,
\begin{equation}
\begin{split}
\widetilde{KE}_{\downarrow}=\frac{1}{N}\sum_{\textbf{k}\textbf{q}} \epsilon_{\textbf{k}-\frac{\textbf{q}}{2}} |v_{\textbf{k}\textbf{q}}|^2 \langle S^-S^+  \rangle _\textbf{q}    - \epsilon_{\textbf{k}'-\frac{\textbf{q}}{2}} |v_{\textbf{k}'\textbf{q}}|^2.
\end{split}
\label{EDNKE}
\end{equation}
The kinetic energy of the magnon takes the following form,
\begin{equation}
\begin{split}
\widetilde{KE}_m =\frac{1}{N}\sum_{\textbf{k}\textbf{q}} \omega_\textbf{q} |v_{\textbf{k}\textbf{q}}|^2 \langle S^-S^+  \rangle _\textbf{q}^2  - \omega_\textbf{q} |v_{\textbf{k}'\textbf{q}}|^2 \langle S^-S^+  \rangle _\textbf{q}^2.
\end{split}
\label{EMAGKE}
\end{equation}
The interaction energy can be written as
\begin{equation}
\begin{split}
\widetilde E_I =\frac{1}{N} \sum_{\textbf{k}\textbf{l}\textbf{q}\textbf{p}}  V_{\textbf{k}\textbf{l}\textbf{q}\textbf{p}}v_{\textbf{l}\textbf{p}} u_{\textbf{l}\textbf{p}}^* v_{\textbf{k}\textbf{q}}^* u_{k\textbf{q}}  {\langle S^-S^+ \rangle_\textbf{q}} {\langle S^- S^+ \rangle_\textbf{p}} \\
- 2\sum_{\textbf{l}\textbf{p}}V_{\textbf{k}'\textbf{l}\textbf{q}\textbf{p}}v_{\textbf{l}\textbf{p}} u_{\textbf{l}\textbf{p}}^* v_{\textbf{k}'\textbf{q}}^* u_{\textbf{k}'\textbf{q}}{\langle S^-S^+ \rangle_\textbf{q}} {\langle S^- S^+ \rangle_\textbf{p}}.
\label{EINTEN1}
\end{split}
\end{equation}
From Eq.\ \ref{deltakq}, we can show that
\begin{equation}
\Delta_{\textbf{k}'\textbf{q}}=-\sum_{\textbf{l}\textbf{p}} V_{\textbf{k}'\textbf{l}\textbf{p}\textbf{q}}v_{\textbf{l}\textbf{p}} u_{\textbf{l}\textbf{p}}^* \sqrt{\langle S^-S^+ \rangle_\textbf{q} \langle S^-S^+ \rangle_\textbf{p}}.
\end{equation}
We use this relation in the right hand side of Eq.\ \ref{EINTEN1}. The second term now gives $+ 2  v_{\textbf{k}'\textbf{q}}^* u_{\textbf{k}'\textbf{q}} \sqrt {\langle S^-S^+ \rangle_\textbf{q}} \Delta_{\textbf{k}'\textbf{q}}$, which, using Eq.\ \ref{sin2theta} gives $\Delta_{\textbf{k}'\textbf{q}}^2/E_{\textbf{k}'\textbf{q}}$.
Then,
\begin{equation}
\begin{split}
\widetilde E_I =\frac{1}{N} \sum_{\textbf{k}\textbf{l}\textbf{q}\textbf{p}}  V_{\textbf{k}\textbf{l}\textbf{q}\textbf{p}}v_{\textbf{l}\textbf{p}} u_{\textbf{l}\textbf{p}}^* v_{\textbf{k}\textbf{q}}^* u_{\textbf{k}\textbf{q}}  {\langle S^-S^+ \rangle_\textbf{q}} {\langle S^- S^+ \rangle_\textbf{p}} +\frac{\Delta_{\textbf{k}'\textbf{q}}^2}{E_{\textbf{k}'\textbf{q}}} .
\label{EINTEN}
\end{split}
\end{equation}
Combining Eqs. (\ref{EUPKE},\ref{EDNKE},\ref{EMAGKE},\ref{EINTEN}),
we get the result that we will use in DOS calculation,
\begin{equation}\begin{split}
&\langle\widetilde{\psi} | H | \widetilde{\psi} \rangle - \langle \psi | H | \psi \rangle = \epsilon_{\textbf{k}'+\frac{\textbf{q}}{2}}+\frac{\Delta_{\textbf{k}'\textbf{q}}^2}{E_{\textbf{k}'\textbf{q}}} \\
&- (\epsilon_{\textbf{k}'+\frac{\textbf{q}}{2}}  + \epsilon_{\textbf{k}'-\frac{\textbf{q}}{2}}   + \omega_\textbf{q}  \langle S^-S^+  \rangle _\textbf{q}) |v_{\textbf{k}'\textbf{q}}|^2   \langle S^-S^+  \rangle _\textbf{q}
\end{split}
\end{equation}

\end{document}